\documentclass[11pt,a4paper]{article}
\usepackage{alltt}
\usepackage{graphicx,subfigure} 
\usepackage[super,numbers,sort&compress]{natbib} 
\usepackage{authblk}
\usepackage{natmove}
\usepackage{xfrac} 
\usepackage{anysize} 
\usepackage{setspace} 
\usepackage{amsmath} 
\usepackage{amssymb} 
\usepackage{float} 
\usepackage{tabu} 
\usepackage{booktabs} 
\usepackage[pdfborder={0 0 0}]{hyperref} 
\usepackage[small,indent]{caption} 

\pdfoutput=1

\providecommand{\keywords}[1]{\vspace*{5mm}\noindent\textit{Keywords:} #1}

\renewcommand\arraystretch{1.2}

\graphicspath{{figure/}}

\newcommand{\specialcell}[2][c]{%
  \renewcommand\arraystretch{1}\begin{tabular}[#1]{@{}c@{}}#2\end{tabular}\renewcommand\arraystretch{1.3}}

\title{Variable selection with genetic algorithms using repeated cross-validation of PLS regression models
as fitness measure}
\author[1]{David Kepplinger\thanks{Corresponding author. E-mail address: \href{mailto:d.kepplinger@stat.ubc.ca}{d.kepplinger@stat.ubc.ca}}}
\author[2]{Peter Filzmoser}
\author[2]{Kurt Varmuza}
\affil[1]{Department of Statistics, University of British Columbia, Vancouver, British Columbia, Canada}
\affil[2]{Institute of Statistics and Mathematical Methods in Economics, TU Wien, Vienna 1040, Austria}
\date{}

\begin{document}
\maketitle

\begin{abstract}
Genetic algorithms are a widely used method in chemometrics for extracting variable subsets with
high prediction power. Most fitness measures used by these genetic algorithms are based on the ordinary least-squares
fit of the resulting model to the entire data or a subset thereof. Due to multicollinearity, partial
least squares regression is often more appropriate, but rarely considered in genetic algorithms due to the
additional cost for estimating the optimal number of components.

We introduce two novel fitness measures for genetic algorithms, explicitly designed to estimate
the internal prediction performance of partial least squares regression models built from the variable subsets. Both measures
estimate the optimal number of components using cross-validation and subsequently estimate the
prediction performance by predicting the response of observations not included in model-fitting.
This is repeated multiple times to estimate the measures' variations due to different random splits. Moreover, one measure
was optimized for speed and more accurate estimation of the prediction performance for observations not included during variable selection.
This leads to variable subsets with high internal and external prediction power.

Results on high-dimensional chemical-analytical data show that the variable subsets acquired by this approach have competitive
internal prediction power and superior external prediction power compared to variable subsets extracted
with other fitness measures.

\end{abstract}

%
%

\section{Introduction}
Building models from chemical-analytical data suitable for predicting future observations is challenging
and often requires the reduction of dimensionality.
In lots of situations, too many variables can dramatically increase the noise in the data and thereby
decrease the descriptive and predictive power of a model. However, deciding which variables carry most of
the relevant information highly depends on the problem. Variables that can represent the data very
well may be less important for predicting future observations and vice versa. Thus, most
variable selection procedures are designed for specific problem settings and data structures.

Predicting future observations is a prevalent use case for models of chemical-analytical data.
Quantitative structure-property relationship models (QSPR), for instance, relate structural data
(molecular descriptors) to properties of the compounds. Predicting the properties of compounds when
only the molecular descriptors are known is an important use of these models. QSPR data often
comprises a huge number of molecular descriptors but the property is known only for a few compounds. As
many of these molecular descriptors are also highly correlated, PLS regression models\cite{Varmuza2009,Wold2001}
are frequently used to fit the model to the data. Therefore, in order to fully specify the PLS regression model, the
complexity of the PLS model must be estimated. This poses the necessity of estimating the number
of PLS components and finding the molecular descriptors most important to predicting the property at question
for other compounds simultaneously.

Once the use of the model and the structure of the data is known, a validation criterion that can assess
the quality of the model with respect to this intended use must be specified. The validation criteria
discussed in this paper have the sole purpose of finding variable subsets with high prediction power.
The need to estimate the number of PLS components complicates the validation even further. Repeated
double cross-validation (rdCV)\cite{Filzmoser2009} is a useful method for validating the prediction
power of PLS regression models. As discussed by \citet{Gramatica2014}, internal and external prediction
power have to be differentiated as well as considered when choosing the validation criterion. Due to
its favorable properties, rdCV will be used as reference for the internal prediction performance of
PLS models throughout this paper.

Many different strategies for variable selection
have been intensively studied in the chemometrics literature. Simple methods for variable
selection are based on the empirical correlation coefficient
between the covariates and the response variable. Another class of methods is stepwise regression.
A single variable is added/removed to/from the model such
that a previously defined model performance criterion is optimized in each step. As ordinary least-squares
regression estimates are used for most stepwise regression approaches, multicollinearity and too many
covariates are a strong limitation for these methods\cite{Gauchi2001}. A comprehensive listing
of these simple variable selection methods is given in \citet{varmuza2013}. Recently, an enhanced version
of well-known sequential replacement methods was published\cite{Grisoni2014}, which gives very promising results for
QSPR models and is also quite fast to compute. Other methods directly use PLS regression models to search for
variable subsets. One form of these methods is based on significance tests, for instance by iteratively removing variables
that are insignificant according to a t-test on the estimated coefficients\cite{Westad2000}.

Genetic algorithms (GA) are another prevalent class of methods used for variable selection in chemometrics.
\cite{Niazi2012} As GAs are highly adaptable, different forms were used in research in the past. However,
all methods share the need for an easy to compute internal fitness measure. As this fitness
measure has to be calculated for many different models, its computational speed is of high priority.
Too complex validation criteria take long time to compute and therefore simple, fit-based criteria
where the regression estimates are only calculated once for the entire data are dominant in the literature.
The mature GA in \citet{gramatica2007} uses the coefficient of determination ($R^2$), a fit-based criterion,
calculated from the leave-one-out (LOO) residuals obtained from an OLS regression model. These LOO residuals
can be easily calculated from an OLS regression model that is once fit to the entire (training) data.
However, as multicollinearity is a big issue for OLS regression, the risk is that variables that are highly
correlated with each other are removed before employing the GA.
Predictive abilities of PLS regression models have already been considered in genetic algorithms.\cite{leardi1998}
However, to simplify computation, the number of optimal PLS components was estimated only once using all
variables, and the chosen criterion to be optimized was the cross-validated explained variance ($Q^2$). A similar
approach\cite{broadhurst1997} was to fit a PLS regression model to a fixed training set and validate
the prediction power on a fixed model test set within the GA. Both methods do not account for
inherent variation between different cross-validation segmentation as well as splits into training and
validation sets. In this paper we propose new measures of prediction power to be
used within GAs that take the variation between different splits of the data into consideration.

We will emphasize the need of accurate estimation of the prediction power \textit{within}
the genetic algorithm. First, the considered validation criteria are discussed in detail. What follows
is a short description of the genetic algorithm used for the comparison of the validation criteria.
After comparison of the variable subsets extracted by using the different validation criteria in the GA,
the best performing models will be put into relation with variable subsets proposed in previous
papers.

%
%

\section{Validation}
To compare two variable subsets, the purpose of the resulting model and the properties
of the data must be taken into account. In chemometrics, the number of covariates is often a multitude
of the sample size and many of these covariates can be highly correlated. Also, as prediction
power of the model is of major interest, the complexity of the model has to be optimized. Partial
least squares (PLS) regression was designed to cope with these issues. Both, the covariates
$\mathbf{X} \in \mathbb{R}^{N \times p}$ and the response $\mathbf{y} \in \mathbb{R}^{N}$ are assumed
to be generated by the same latent variables (components).
With the number of PLS components $A$, the complexity of the model can be adjusted to avoid overfitting
and increase prediction power. Due to these favorable properties, PLS regression is a prevalent tool in
chemometrics\cite{Wold2001,wehrens2011}.

\subsection{Estimating the Number of PLS Components}
As the number of PLS components determines the complexity of the resulting model and is generally not
known beforehand, it must be chosen sensibly. Too few components will result in an inappropriate
model that is not able to fit the essential structure of the data, while too many components will result
in a model overfitted to the available samples. Considering this, cross-validation (CV) is
a widely used method for estimating the optimal number of components.

Cross-validation works by repeatedly fitting the model to a portion of the data and validating the prediction
performance on the other part of the data. First, the data is split randomly into $K$ almost equally sized
segments. $K - 1$ segments are then used to fit PLS models with one to the maximum of $A_\mathrm{max}$ components.
These fitted PLS models are used to predict the values from the left out segment (validation set). When this
procedure has been repeated $K$ times, such that each segment takes the role of the validation set exactly once,
$A_\text{max}$ predicted values $\hat{y}_i^{(a)}$ are available for every observation $y_i$.
Let $\mathcal{J}_k$ denote the indices of observations belonging to segment $k$, $k = 1, \dotsc, K$.
The mean squared error of prediction $\text{MSEP}_a$ can be calculated for each number of components
$a$ according to
\begin{equation}
\text{MSEP}_{a} = \frac{1}{K} \sum_{k = 1}^K \frac{1}{| \mathcal{J}_k |} \sum_{j \in \mathcal{J}_k}
        \left( y_j - \hat{y}_j^{(a)} \right)^2
,
\quad a = 1, \dotsc, A_\text{max}.
\end{equation}

$\text{MSEP}_a$ guides the search for the optimal number of components. $A_\text{opt}$ can now be chosen as
the number of components that results in the minimal $\text{MSEP}_a$ or by other
heuristics. An often used strategy is taking the smallest number of components that results in a $\text{MSEP}_a$
that is still less than the minimal $\text{MSEP}_a$ plus one standard error  (the one standard
error rule as described in \citet[Figure~3.7]{hastie2009}), given by
\begin{equation}
A_\text{opt} := \min\left\{ a : \text{MSEP}_{a} \leq \text{MSEP}_m + \frac{\text{SE}_m}{\sqrt{K}} \right\}
\quad
\text{with }
m = \operatorname*{arg\,min}_a \text{MSEP}_a,
\end{equation}
whereas the standard error $SE_a$ of the MSEP for the model with $a$ components is defined as
\begin{equation}
\text{SE}_{a} = \sqrt{\frac{1}{K - 1} \sum_{k = 1}^K \left(\text{MSEP}_a -
        \frac{1}{| \mathcal{J}_k |} \sum_{j \in \mathcal{J}_k} \left( y_j - \hat{y}_j^{(a)} \right)^2
    \right)^2}.
\end{equation}

Once the optimal number of components is estimated, validation of the model for the desired use is mandatory.
For validating the prediction power of the model, an independent set of observations is desirable but seldom extant.
To circumvent the need of an independent set of observations, different strategies have been proposed
in the literature. Measures acquired during the CV for model-calibration are too optimistic,\cite{Filzmoser2009}
so other measures to validate the model have to be used.

The easiest and fastest way to validate any model is to quantify how well the model fits the data. No
independent data set is needed for validation and the computational burden is very low as only one
model with $A_\text{opt}$ number of components has to be fit to the data.
Prominent methods to assess the model-fit are well-known information criteria like Akaike's information
criterion (AIC) and the Bayes information criterion (BIC) or the coefficient of determination ($R^2$).
Because $R^2$ gets larger the more covariates are used in
the model, the adjusted $R^2$ is usually preferred when models with different numbers of predictors
are to be compared as it penalizes greater numbers of predictors. However, if prediction power of the
model is of primary interest, fit-based criteria are not able to quantify the desired properties.

\subsection{Repeated Double Cross-Validation}
Especially when the number of available observations in the data set is small, which is often the case
for QSPR models,\cite{varmuza2013} resampling methods are the only way to estimate the prediction
performance of the model.\cite{Gramatica2014}

\begin{figure}[t]
{\centering
    \includegraphics[width=130mm]{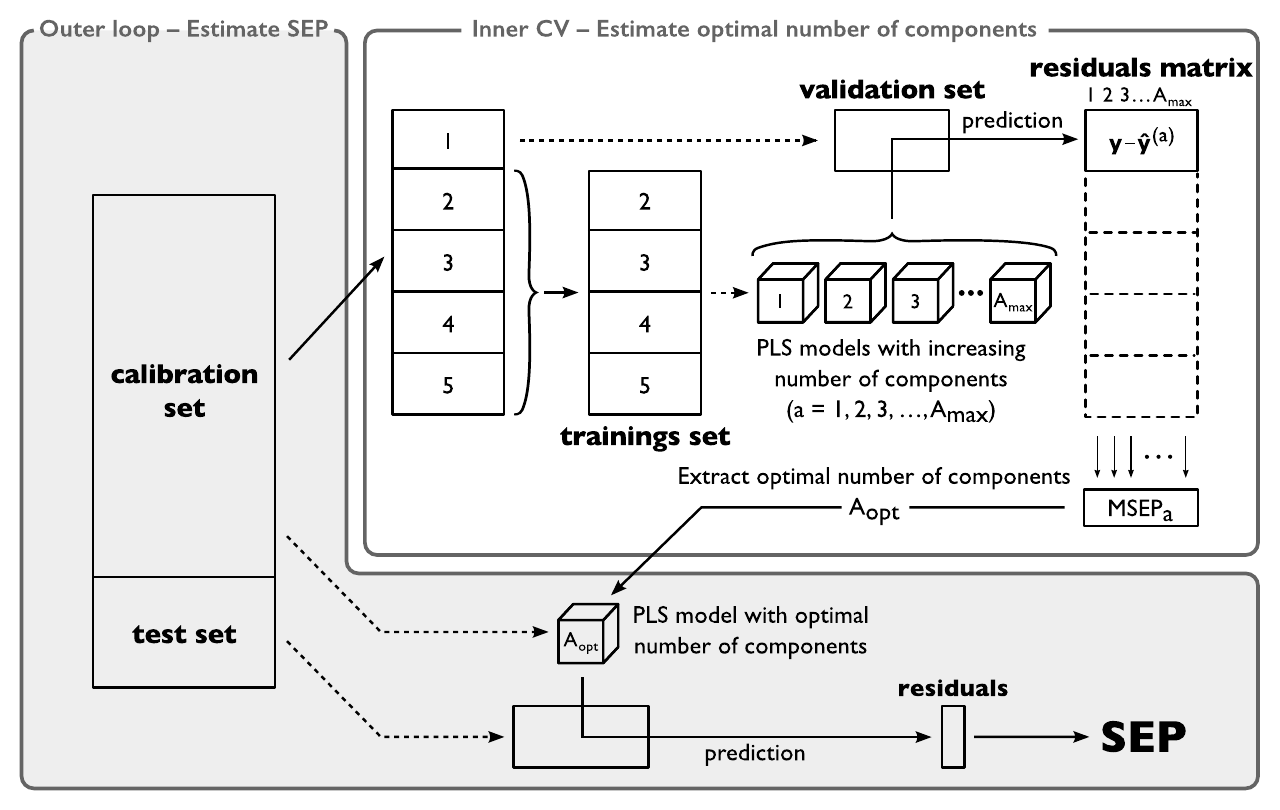}

}
\caption[Scheme of a single replication for simple repeated cross-validation]{Scheme of a
\textit{single} replication for simple repeated cross-validation (modified from
\citet[p.~3]{varmuza2013})\label{fig:srcv-scheme}}
\end{figure}

\citet{Filzmoser2009} proposed the repeated double cross-validation (rdCV) strategy, an extension
to double cross-validation,\cite{Baumann2014}  as reliable estimate
for the prediction performance of PLS regression models. Similar to double cross-validation it
consists of an outer CV loop and an inner CV
loop. In the outer CV loop, the data are split into $S$ segments. The calibration set consist of $S - 1$ segments
and the remaining segment is used as test set (see Figure~\ref{fig:srcv-scheme}). Using the calibration
set, the optimal number of components $A_\text{opt}$ is estimated in the inner CV loop. The model
with $A_\text{opt}$ components is then fit to this calibration set. The fitted model is finally used
to predict the response in the test set.
This is repeated such that each of the $S$ outer segments is used as test set exactly once,
hence for every observation $y_i$ a predicted value $\hat{y}_i$ is available and the information
of $y_i$ and $\mathbf{x}_i$ is not used for this prediction.
With these predictions the Standard Error of Prediction (SEP) can be calculated as
\begin{equation}
\begin{split}
\text{SEP} &= \sqrt{ \frac{1}{N - 1} \sum_{i = 1}^{N} \left( y_i - \hat{y}_i - \text{bias} \right)^2 } \\
\text{where } \text{bias} &= \frac{1}{N} \sum_{i = 1}^{N} \left( y_i - \hat{y}_i \right).
\end{split}
\end{equation}
A similar concept and also often used for model assessment is the Root Mean Squared Error of Prediction (RMSEP).
It is given by
\begin{equation}
\text{RMSEP} = \sqrt{\frac{1}{N} \sum_{i = 1}^N (y_i - \hat{y_i})^2}
\label{eq:rmse}
\end{equation}
and can be expressed in terms of the standard error of prediction
$\text{RMSEP} = \sqrt{\frac{N - 1}{N} \text{SEP}^2 + \text{bias}^2}$.

Although double cross-validation gives a reliable estimate of the prediction error,\cite{Baumann2014}
the segmentations in both the outer and the inner CV loop is random and hence the estimated SEP is a random quantity
as well. To get viable information about it's variance, a single value is insufficient.
In rdCV, the procedure is therefore repeated with different splits in the outer as well as in the inner CV loop. A final
estimate for the prediction performance is then the arithmetic mean of all SEP replications.

The outer and inner CV loop make the rdCV estimate computationally expensive, which can be a huge handicap
when many different models need to be validated. Also, in case of data sets with only a small number
of observations, the single segments and thus the test set in rdCV contains only a few observations
($N_\text{test}$ is very small). As this can lead to inappropriate estimates and long computation times, a
simplified version is introduced.

\subsection{Simple Repeated Cross-Validation}

Simple repeated cross-validation (srCV) is a simplified and hence faster version of rdCV.
In contrast to rdCV, the outer loop is not a cross-validation loop, but the data set is only
split at random into a calibration set and a test set. The calibration set is again used to estimate
the optimal number of components according to the scheme described in the above section. Likewise, the resulting model
with $A_\text{opt}$ components is then fit to the complete calibration set. This fitted model is subsequently
used to predict the responses from the test set. Finally, the standard deviation of the resulting residuals is used as an estimate for the SEP.

As with rdCV, the estimate varies with different splits into calibration and test sets. To assess the
variability and to get a more reliable estimate for the SEP, the procedure is repeated multiple times
with different splits into calibration and test set. The arithmetic mean of all replicated
SEP values is taken as the final estimate of the prediction power.

The simplification engenders another advantage over rdCV. The size of the test set is not fixed
to one segment of the outer cross-validation loop, but can be adjusted to include an arbitrary number
of observations.

\subsection{External Validation}
According to \citet{Gramatica2014}, external and internal validation methods must be differentiated.
Although above methods use one part of the data for fitting the model(s) and the other part for
assessing the prediction power, due to the replications, information from the entire data is used
nevertheless. Therefore, they are considered as internal validation criteria and tend to overestimate
the prediction power. To estimate the prediction power for totally new -- external -- observations,
the model has to be validated with data that was never used during model selection. This is particularly
challenging when only few observations are available.

%
%

\section{Genetic Algorithm}
To compare the different validation criteria in the variable selection setting, a simple genetic algorithm
(GA) was used. Genetic algorithms\cite{Goldberg1989} are a very general class of methods for
finding a global optimum in a large search space with no assumptions on the objective function, by
combining a guided and a random search, mimicking strategies from evolution. To reduce the size of the
search space and therefore the time complexity, only variable subsets within a minimum and maximum number
of variables are considered. For this, a GA following the scheme described in \citet{leardi2007}
that supports all aforementioned validation criteria was implemented for the statistical software
environment R\cite{rsoftware} and is available as package \textit{gaselect} on CRAN.\footnote{\url{http://cran.r-project.org/package=gaselect}}

Genetic algorithm terminology highly draws from genetics. Points in the search space are denominated as
\textit{chromosomes}
and they are defined by their \textit{genes}. Every chromosome has an associated \textit{fitness} value
that forms the objective function the GA optimizes. In GAs for variable selection, every possible
variable subset is represented by a chromosome and the fitness of a chromosome is calculated with
one of the validation criteria previously defined.

\begin{figure}[tb]
{\centering \includegraphics[width=80mm]{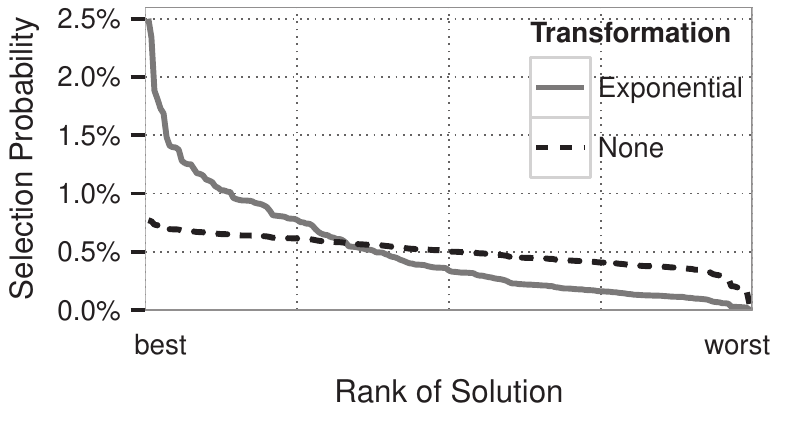}

}
\caption[Impact of transforming the selection probabilities with the exponential function on exemplary fitness values]{Impact of transforming the selection probabilities with the exponential function on exemplary fitness values. The left-most chromosome has the highest fitness and the right-most chromosome the lowest fitness. Without transformation, the selection probabilities for most chromosomes are similar, causing the algorithm to converge to the minimum at a slower rate.}\label{fig:fitness-scaling}
\end{figure}

The search starts by selecting a large number of different chromosomes and evaluating their \textit{fitness}.
Starting from this initial \textit{generation}, evolution is imitated by randomly combining two
chromosomes (\textit{mating}) to form two offsprings and randomly adding/removing variables (\textit{mutation})
to/from these offsprings. Like in evolution, chromosomes with higher fitness have a greater chance
to be selected for mating and thus have a higher chance to produce offsprings. For selecting the chromosomes,
the fitness $f(c)$ of chromosome $c$ is standardized by $f^\ast(c) = \frac{f(c) - \overline{f(c)}}{SD(f(c))}$
before the selection probabilities are assigned. As extremely bad chromosomes result in almost equal
selection probabilities for all solutions, even for very good ones, the scaled fitness $f^\ast(c)$ is
optionally transformed with the exponential function $f^{\ast\ast}(c) = exp\left( f^\ast(c) \right)$.
Figure~\ref{fig:fitness-scaling} shows the difference between $f^\ast(c)$ and $f^{\ast\ast}(c)$ in
the selection probabilities of chromosomes in a generation with only few bad solutions present.

Once two chromosomes are selected, the mating process takes place. The two most common ways
to combine two chromosomes are \textit{single crossover} or \textit{uniform crossover}.\cite{leardi2007}
For single crossover, one gene is randomly selected as splitting point. The first offspring
is formed by the genes left and including the selected gene from the first parent and the genes
to the right of the selected
gene from the second parent. The second offspring is formed in a similar fashion by exchanging the two
parents' roles. Uniform crossover randomly selects an arbitrary number of genes. The selected genes
from the first parent and the non-selected genes from the second parent shape the first offspring. Again,
by exchanging the role of the two parents, the second offspring is created.

\begin{figure}[tb]
{\centering \includegraphics[width=80mm]{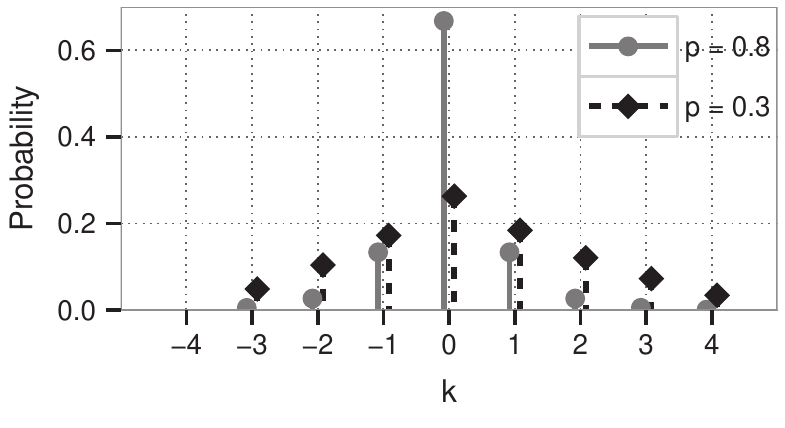}

}
\caption[Probability mass function ]{Probability mass function $g(k)$ of a truncated double geometric distribution with $l = -3$ and $u = 4$.}\label{fig:fig-ddtgeom}
\end{figure}

To give the GA a chance to elude local optima, the offspring's genes can mutate with a small probability.
The number of mutated genes follows a \textit{double truncated geometric distribution} with probability mass
function shown in Figure~\ref{fig:fig-ddtgeom} and given by
\begin{equation}
g(k) = \frac{p \left( (1-p)^{2 + k + l - 2 \max(l, k - u)} - (1 - p)^{k + l - 2 \min(0, k)} \right)}
{(p - 2) (1 - (1 - p)^{1 + u}) (p - 1 + (1 - p)^{l})}
\quad
k \in [l, u]
\end{equation}
where $1 - p$ is the mutation probability, $u$ ($ > 0$) is the upper truncation point and $l$ ($< 0$) is the lower truncation
point, hence the support of the distribution is between $l$ and $u$. The distribution needs to be truncated
to guarantee that the number of variables stays between the set limits. This distribution arises as the
difference between two truncated geometric distributed random variables,\cite{Thomasson1968} one with truncation point $u$ and
the other one with truncation point $-l$. Thus, it is improbable that many genes mutate and the number
of variables in a subset will never be outside the specified bounds.

The fitness of the new offspring has to be evaluated with a selected validation criterion.
However, an offspring will not be accepted if it is a certain degree worse than the parent with the lower
fitness, thus very bad combinations or mutations are discarded right away. The offspring will also
be rejected if it is a duplicate of another chromosome in the offspring-generation. Consequently,
a high level of diversity in the new generation is maintained and the average fitness of the population
increases.

If an offspring is extraordinary fit it will join the \textit{elite}.
Elite chromosomes have the chance for mating in every generation, but if the group gets too large,
the worst chromosome from the elite will be discarded.

When there are as many offsprings as there were chromosomes in the previous generation, these offsprings
form the new generation and may produce offsprings themselves. The GA will stop once a predefined
number of generations are generated.
More details about genetic algorithms can be found in previously published literature.\cite{Goldberg1989,wehrens2011,leardi2007}

The computationally most expensive step is to evaluate a chromosome's fitness. To mitigate the computational
burden, genetic algorithms can be parallelized to distribute the work load to multiple processing units.
One way is to split the entire population into a number of completely independent subpopulations of
smaller size. This could be extended to allow some chromosomes to switch the subpopulation at certain
times. Another approach, implemented in the GA utilized to compare the validation criteria in this work,
is to use one big population and distribute the mating, mutation, and evaluation step to multiple nodes.
Every node produces a fixed number of offsprings according to the above steps, without need
for interaction between the nodes and therefore almost no overhead caused by parallel computation.

We have implemented the SIMPLS algorithm\cite{jong1993} to calculate the PLS regression estimates.
As it is performed numerous times, we aggressively optimized the algorithm for single response models.
As pointed out by \citet{Faber2008}, the SIMPLS algorithm is numerically quite unstable due to the
orthogonality requirement for the loadings. To mitigate this problem, we employ the modified
Gram-Schmidt (MGS) orthogonalization process, but avoided the recommended reorthogonalization
due to the performance penalty. Thus, the algorithm may give
inaccurate results for models with an extraordinary large number of components.

\section{Applications}
Performance of the validation criteria for variable selection is compared by checking the internal
and external prediction power of the resulting variable subsets. Two real-life QSPR data sets will be used for
this analysis.

Additionally to the rdCV and the srCV, two fit-based validation criteria are also considered in the
following. The most naive approach is to use the BIC value
\begin{align}
\text{BIC} &= N \log(\text{RSS} / N) + p \log(N) \\
\text{where } \text{RSS} &= \sum_{i = 1}^{N} \left( y_i - \hat{y}_i \right)^2 \nonumber
\label{eq:bic}
\end{align}
obtained from an ordinary least squares fit
to the entire data (denoted by $\text{BIC}_\text{OLS}$). This method is of course only applicable when
the number of variables in the
considered model is less than the number of observations and no pair of selected variables is perfectly
collinear. In the examples considered in this paper, we always try to find models with only a few variables and
therefore the only problem is multicollinearity. When the OLS estimate for a model can not be computed,
the model is not considered further in the GA. The other fit-based method considered is the BIC value
obtained from a PLS regression model, also fit to the entire data (denoted by $\text{BIC}_\text{PLS}$).
The number of PLS components is estimated with the cross-validation procedure described above. We
include these two methods to compare our proposed validation criteria to commonly used ones.

Performance of the new internal fitness measures is compared to the established fit-based fitness
measures by employing them in the GA applied to two high-dimensional QSPR data sets.
Both data sets have more variables than observations and multiple highly correlated variables.
The validation criteria will be compared to each other and to results obtained in previously published papers
\cite{gramatica2007,varmuza2013} concerned with these data sets.

\subsection{Data sets}
\paragraph{KOC data set.} Data of the soil sorption coefficient, normalized on organic carbon
($\mathrm{K_\mathrm{oc}}$) for $N=643$ heterogeneous organic compounds\cite{gramatica2007} with $p=1266$
molecular descriptors. Because the compounds are very heterogeneous, it is difficult to find variable
subsets that give a good prediction for all different kinds of compounds.

\paragraph{PAC data set.} The GC retention index for $N=209$ polycyclic aromatic compounds (PAC) with
$p=2688$ molecular descriptors.\cite{varmuza2013} This data set is particularly demanding for variable
selection algorithms due to the very high number of variables, multicollinearity and comparatively
few observations. In \citet{varmuza2013} different variable selection procedures were used to find
good variable subsets for prediction and their results will be compared to the results in this work.

\subsection{Results}
The GA described in the previous section is used with the different validation criteria to search for
suitable variable subsets. For both data sets, the GA was employed with a population size of 4000
chromosomes and generated 300 generations. The three validation criteria using PLS regression models
(rdCV, srCV, and $\text{BIC}_\text{PLS}$) were configured to estimate the optimal number of components
with 10 CV segments (inner loop). The rdCV criterion was used with four outer CV segments, while the
srCV criterion used 60 percent of the data for calibration and the other 40 percent as test set. Both
criteria were employed with 30 replications. Due to the larger number of variables and therefore larger
search-space, the mutation probability for the GA applied to the PAC data set was set to $0.5$ percent,
compared to $0.25$ percent for the GA applied to the KOC data set. For the PAC data set, the GA was
searching for variable subsets with 3 to 30 variables, whereas for the KOC data set, the number of
variables was limited to 10.

Numerous variable subsets of the PAC data set give extremely large SEP values. The average fitness of the
initial population is more than 4.5 times as large as the fitness of the final solution. Variable subsets
from the KOC data set have similar properties, albeit less severe. As this has a significant influence
on the chromosome selection during the GA, the fitness values have been transformed
as outlined in the previous section.

\begin{figure}[tb]
{\centering \includegraphics[width=140mm]{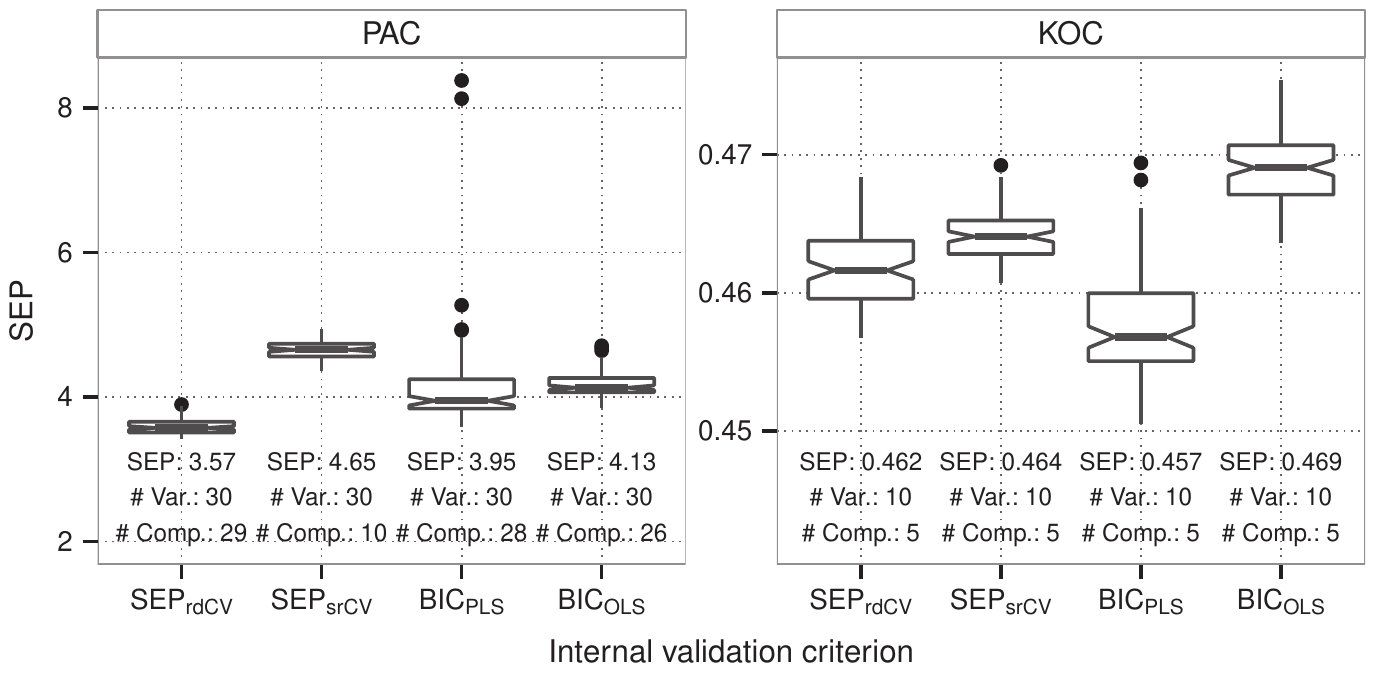}

}
\caption[Distribution of the estimated SEP using the rdCV implementation provided by the R package \texttt{chemometrics}\cite{rpkgchemometrics}
for variable subsets found with the GA using different internal validation criteria]{Distribution of the estimated SEP
using the rdCV implementation provided by the R package \texttt{chemometrics}\cite{rpkgchemometrics}
for variable subsets found with the GA using different internal validation criteria.}\label{fig:subsets-internal}
\end{figure}

Internal prediction power was verified for the ten top ranked
variable subsets (according to the used validation criterion) with the rdCV implementation provided in the R package \texttt{chemometrics}
\cite{rpkgchemometrics}. The procedure was run with 50 replications, 10 inner and 4 outer CV segments
using the \textit{kernelpls} algorithm for PLS regression (``algorithm 1'' in \citet{Dayal1997}) as
this algorithm is numerically more stable than SIMPLS. For rdCV and srCV, the results of the R procedure
were very similar to the internal prediction power estimated by the GA.

Of these ten top ranked variable subsets, the SEP of the best subset according to the verified internal prediction power
for each data set and validation criterion is shown as boxplot in Figure~\ref{fig:subsets-internal}.
Additionally, the size of the variable subset as well as the estimated optimal number of components
is given in the graph. The variable subset for PAC found using rdCV as
validation criterion gives the best results in terms of SEP. However, it is interesting that the two fit-based
validation criteria result in variable subsets with a lower SEP than the srCV criterion. Moreover,
the variability of the SEP of variable subsets extracted with the fit-based criteria is much higher
than for the other two criteria.
The notches of the boxes do not overlap, which is an indication that the differences of the medians
are significant. All four validation criteria combined with the GA give significantly better variable
subsets than the simple variable selection methods demonstrated in \citet{varmuza2013}. The best (in
terms of rdCV) variable subset in \citet{varmuza2013} leads to a SEP of approximately $5.6$, compared
to the minimum SEP of $3.57$ reported here.

With regard to the internal prediction power of the variable subsets found for the KOC data set,
the fit-based criterion using PLS regression models gives the best result. However, the SEP is only 1 percent
smaller than the SEP of the variable subset found with rdCV and variability of the SEP is quite large.
Such solutions which result in large SEP values for certain CV splits will less likely be considered
by rdCV and srCV, as both criteria use the arithmetic mean over all different splits and are therefore
more affected by these extreme values. Therefore, the internal prediction power of the variable subsets
obtained with rdCV is very competitive.

A major weakness of the GA is immediately apparent from the found variable subsets: almost all variable
subsets have the maximum allowed number of variables. The main reason for this is the very primitive mating
procedure implemented in the GA. We do not take the number of variables of the parents into account
before the mating takes place. Therefore, child chromosomes can have up to double the number of allowed
variables. The excessive variables are randomly removed to constrain the number of variables to the specified
range. Thus, a reasonable enhancement to the GA would be to take the number of variables of the parents
into account prior to mating.

\begin{figure}[tb]
{\centering \includegraphics[width=140mm]{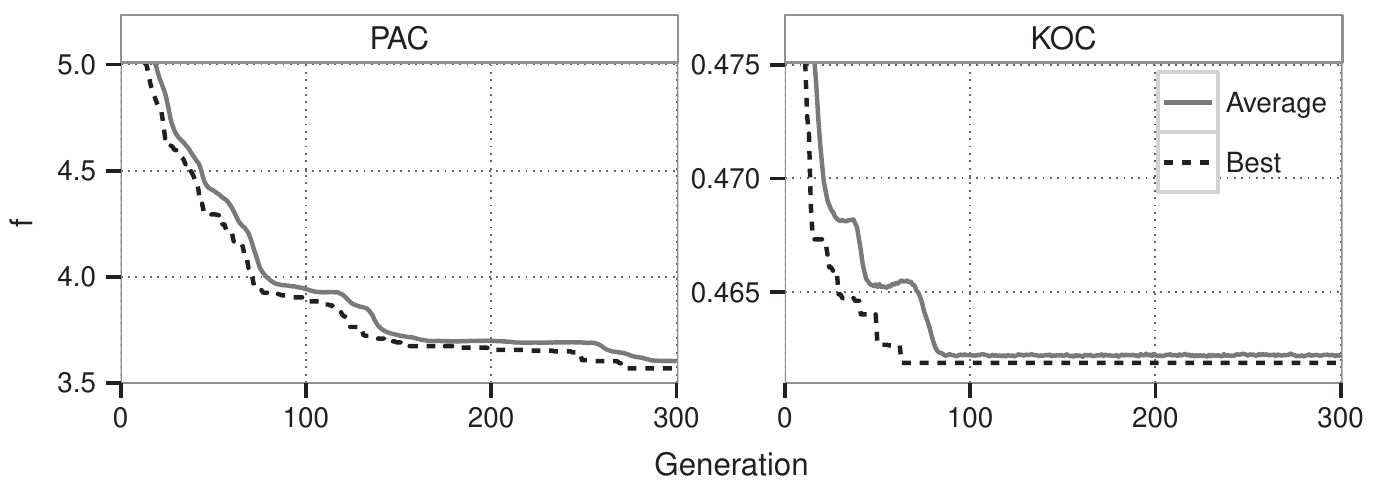}

}
\caption[Progression of the average fitness and the fitness of the best chromosome over generations]{Progression
of the average fitness and the fitness of the best chromosome over generations. Results taken from the GA run that
resulted in the variable subset with the lowest SEP.}\label{fig:fitness-history}
\end{figure}

Visualization of the average and best SEP during the run of the genetic algorithm in Figure~\ref{fig:fitness-history}
supports the statement from above, that the average fitness increases over generations. The GA applied to the
KOC data set could have been stopped earlier, as the fitness did not increase at all after the first 100
generations. However, unlike many other GAs, the utilized GA does not use any form of stopping criterion when the fitness
is not improving anymore. Additionally, the graph shows that there exist multitudinous variable
subsets which yield an extremely large SEP, underlining the need for transformation of the fitness $f(c)$
to $f^{\ast\ast}(c)$ when assigning the selection probabilities.

External prediction power of the variable subsets selected with the different criteria was estimated
by applying the GA to a subset of the data (external training set) and calculating the root mean squared error of prediction
for the predictions of the other part of the data (external validation set). For the PAC data set, 60 percent of the observations
were randomly chosen to form the external training set. To get an estimate of the variability induced by the random
split, we repeated the entire process 10 times. In the case of the KOC data set, we used the same 93 samples
as in the paper by \citet{gramatica2007} for variable selection and the other 550 samples for validation
in order to make the results comparable.

\begin{table}[tb]
\centering
{\footnotesize
\caption{External prediction power of variable subsets extracted by the GA using different validation criteria for the two data sets.}\label{tbl:subsets-external}
\begin{tabu}{clrrrrrrrl}
    \toprule
        \specialcell{Data\\set} & \specialcell{Valid.\\criterion} & \specialcell{No. obj.\\training} & \specialcell{No. obj.\\validation} &
            \specialcell{No.\\var.} & \specialcell{No.\\comp.} & \specialcell{RMSEP\\ext. training} & \specialcell{RMSEP\\ext. validation} &
            \specialcell{RMSEP\\total} & \\
    \midrule
        PAC & $\text{SEP}_\text{rdCV}$ & 125 &  84 & 30 & 13 -- 23 & $3.33 \pm 0.54$ & $8.31 \pm 1.2$ & $5.66 \pm 0.6$ &  \\
        PAC & $\text{SEP}_\text{srCV}$ & 125 &  84 & 30 & 9 -- 17 & $4.17 \pm 0.23$ & $7.33 \pm 0.74$ & $5.37 \pm 0.44$ & * \\
        PAC & $\text{BIC}_\text{PLS}$ & 125 &  84 & 30 & 25 -- 30 & $ 2.8 \pm 0.45$ & $8.92 \pm 0.79$ & $5.83 \pm 0.52$ &  \\
        PAC & $\text{BIC}_\text{OLS}$ & 125 &  84 & 30 & 26 -- 30 & $3.34 \pm 0.74$ & $ 8.9 \pm 1.7$ & $5.81 \pm 0.95$ &  \\
    \midrule
        KOC & $\text{SEP}_\text{rdCV}$ &  93 & 550 & 10 &  4 & $0.407$ & $0.519$ & $ 0.5$ &  \\
        KOC & $\text{SEP}_\text{srCV}$ &  93 & 550 & 10 &  4 & $0.374$ & $0.503$ & $0.487$ & * \\
        KOC & $\text{BIC}_\text{PLS}$ &  93 & 550 & 10 &  8 & $0.372$ & $0.592$ & $0.559$ &  \\
        KOC & $\text{BIC}_\text{OLS}$ &  93 & 550 & 10 &  8 & $0.387$ & $0.616$ & $0.581$ &  \\
    \bottomrule
\end{tabu}

\captionsetup{format=plain,indention=0cm,font=footnotesize}
\caption*{The external training and external validation set are randomly chosen for the PAC data set, hence the median and median
absolute deviation calculated from 10 runs are presented. The variable subset with the best external prediction power
for each dataset is highlighted with an asterisk}
}
\end{table}

In terms of external prediction power, the srCV validation criterion outperforms the other two criteria.
As listed in Table~\ref{tbl:subsets-external}, the rdCV criterion and the fit-based methods find variable
subsets with better internal prediction power (\textit{RMSEP ext. training}). However, the variable subsets extracted
with the srCV criterion is superior when predicting observations that were not used during the GA to
find the variable subset (\textit{RMSEP ext. validation}) and when considering the RMSEP for all observations
(\textit{RMSEP total}). Moreover, variable subsets found with srCV tend to result in
simpler PLS regression models with less numbers of components. We can also see that the RMSEP for the
training data is significantly lower than the RMSEP for the validation set with all validation criteria,
hence prediction power of the model is overestimated by all criteria. The simplifications in srCV were initially targeted to speed up
the computations. However, by loosing the restrictions on the size of the test set, the validation
criterion is able to estimate the external prediction power of the model more accurately.

The variable subsets found with srCV using only 93 samples from the KOC data set is
very competitive. It is superior to the variable subsets reported in
\citet{gramatica2007} in terms of RMSEP for external as well as internal observations. Therefore,
the RMSEP for the total data set is almost 10 percent lower than for the best performing subset reported
in their paper ($\text{RMSEP} = 0.532$).

\begin{figure}[tb]
{\centering \includegraphics[width=140mm]{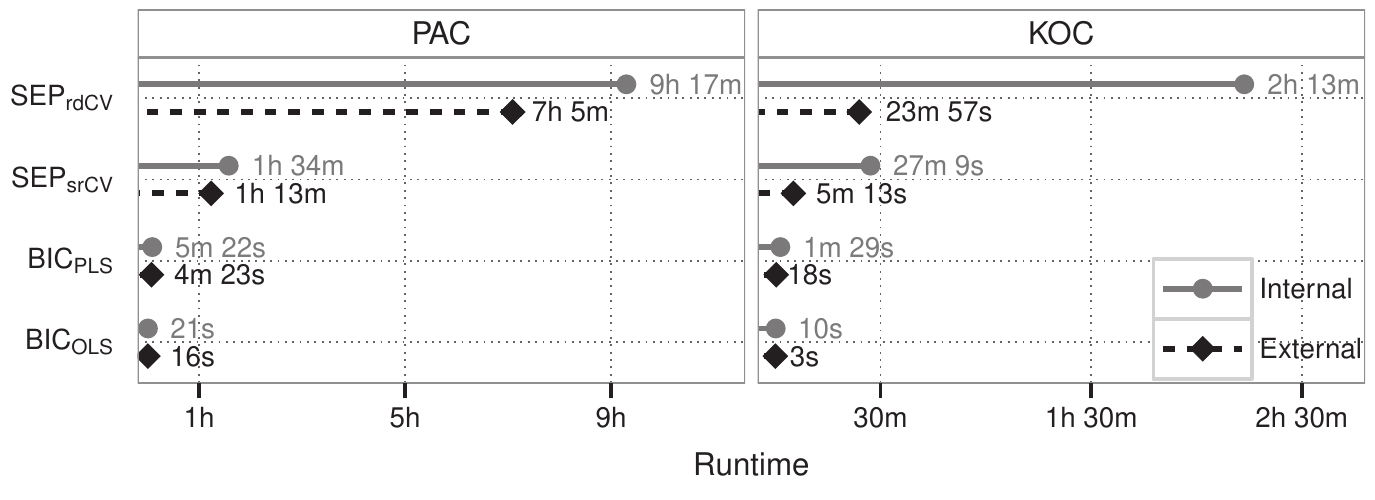}

}
\caption[Runtime of the GA (distributed to 32 threads) with different validation criteria applied to the entire data
\textit{(Internal)} or a subset of the data \textit{(External)}]{Runtime of the GA (distributed to 32 threads) with different
validation criteria applied to the entire data \textit{(Internal)} or a subset of the data \textit{(External)}.}\label{fig:runtimes}
\end{figure}

Currently the major drawback of the proposed validation criteria is the high computation time required. The
computations for each run were distributed to 32 threads and the times recorded (Figure~\ref{fig:runtimes}).
The simplifications in srCV from rdCV
are very effective, decreasing computation time by more than 500 percent compared to rdCV. Of course,
the simple fit based criteria are still significantly faster, but the resulting variable subsets are suboptimal.
Also, as discussed above, the genetic algorithm generated 300 generations regardless of significant
improvements between generations. The current GA implementation leaves room for many more improvements
in terms of fitness of the solution as well as speed. For instance, a more sophisticated GA could
stop the evolution as soon as the fitness does not improve significantly over some generations and
therefore finish faster.

\section{Conclusion}
In summary, the selection of variable subsets from data sets is highly dependent on the purpose of the final model.
By using PLS regression, the models are able to cope with multiple obstacles often observed in chemometrics:
more variables than observations and multicollinearity. The examples given showed that
both validation criteria presented in this paper, repeated double cross-validation and
simple repeated cross-validation, give more realistic estimates of the
internal and external prediction performance of the resulting PLS regression model then other criteria.

The PLS regression models returned by a GA with rdCV and srCV have a very high internal and external
prediction power. Altough the smaller number of variables facilitates interpretation of the resulting model,
interpretability was not the primary goal in the design and selection of the validation criteria. Compared to models previously published
for the examined data sets, both validation criteria perform considerably better. By applying a GA with srCV,
variable subsets with extremely high external prediction power are extracted. The srCV validation
criterion is applicable to a wide range of data sets when variable subsets with high prediction power
are needed.

The genetic algorithm with the repeated double cross-validation, simple repeated cross-validation
measures, and the other presented internal fitness measures are implemented in the R package
\textit{gaselect} and available for download from CRAN\footnote{\url{http://cran.r-project.org/package=gaselect}}.

\section{Acknowledgement}
This work was supported by the Austrian Science Fund (FWF), project 26871-N20.

\bibliographystyle{abbrvnat}
\bibliography{Bibliography_abbr}

\begin{thebibliography}{23}
\providecommand{\natexlab}[1]{#1}
\providecommand{\url}[1]{\texttt{#1}}
\expandafter\ifx\csname urlstyle\endcsname\relax
  \providecommand{\doi}[1]{doi: #1}\else
  \providecommand{\doi}{doi: \begingroup \urlstyle{rm}\Url}\fi

\bibitem[Baumann and Baumann(2014)]{Baumann2014}
D.~Baumann and K.~Baumann.
\newblock Reliable estimation of prediction errors for qsar models under model
  uncertainty using double cross-validation.
\newblock \emph{J. Cheminf.}, 6\penalty0 (1):\penalty0 47, 2014.
\newblock ISSN 1758-2946.

\bibitem[Broadhurst et~al.(1997)Broadhurst, Goodacre, Jones, Rowland, and
  Kell]{broadhurst1997}
D.~Broadhurst, R.~Goodacre, A.~Jones, J.~J. Rowland, and D.~B. Kell.
\newblock Genetic algorithms as a method for variable selection in multiple
  linear regression and partial least squares regression, with applications to
  pyrolysis mass spectrometry.
\newblock \emph{Anal. Chim. Acta}, 348\penalty0 (1--3):\penalty0 71--86, 1997.

\bibitem[Dayal and MacGregor(1997)]{Dayal1997}
B.~S. Dayal and J.~F. MacGregor.
\newblock Improved {PLS} algorithms.
\newblock \emph{J. Chemom.}, 11\penalty0 (1):\penalty0 73--85, 1997.
\newblock ISSN 1099-128X.

\bibitem[de~Jong(1993)]{jong1993}
S.~de~Jong.
\newblock {SIMPLS}: an alternative approach to partial least squares
  regression.
\newblock \emph{Chemom. Intell. Lab. Syst.}, 18:\penalty0 251--263, 1993.

\bibitem[Faber and Ferré(2008)]{Faber2008}
N.~M. Faber and J.~Ferré.
\newblock On the numerical stability of two widely used {PLS} algorithms.
\newblock \emph{J. Chemom.}, 22\penalty0 (2):\penalty0 101--105, 2008.
\newblock ISSN 1099-128X.

\bibitem[Filzmoser and Varmuza(2012)]{rpkgchemometrics}
P.~Filzmoser and K.~Varmuza.
\newblock \emph{chemometrics: Multivariate Statistical Analysis in
  Chemometrics}, 2012.
\newblock {R} package version 1.3.8.

\bibitem[Filzmoser et~al.(2009)Filzmoser, Liebmann, and Varmuza]{Filzmoser2009}
P.~Filzmoser, B.~Liebmann, and K.~Varmuza.
\newblock Repeated double cross validation.
\newblock \emph{J. Chemom.}, 23\penalty0 (4):\penalty0 160--171, 2009.
\newblock ISSN 1099-128X.

\bibitem[Gauchi and Chagnon(2001)]{Gauchi2001}
J.-P. Gauchi and P.~Chagnon.
\newblock Comparison of selection methods of explanatory variables in {PLS}
  regression with application to manufacturing process data.
\newblock \emph{Chemom. Intell. Lab. Syst.}, 58\penalty0 (2):\penalty0 171 --
  193, 2001.
\newblock ISSN 0169-7439.

\bibitem[Goldberg(1989)]{Goldberg1989}
D.~Goldberg.
\newblock \emph{Genetic Algorithms in Search, Optimization, and Machine
  Learning}.
\newblock Artificial Intelligence. Addison-Wesley, Boston, MA, 1989.
\newblock ISBN 9780201157673.

\bibitem[Gramatica(2014)]{Gramatica2014}
P.~Gramatica.
\newblock External evaluation of {QSAR} models, in addition to
  cross-validation: Verification of predictive capability on totally new
  chemicals.
\newblock \emph{Mol. Inf.}, 33\penalty0 (4):\penalty0 311--314, 2014.
\newblock ISSN 1868-1751.

\bibitem[Gramatica et~al.(2007)Gramatica, Giani, and Papa]{gramatica2007}
P.~Gramatica, E.~Giani, and E.~Papa.
\newblock Statistical external validation and consensus modeling: A {QSPR} case
  study for $\mathrm{K_{oc}}$ prediction.
\newblock \emph{J. Mol. Graphics Modell.}, 25\penalty0 (6):\penalty0 755--766,
  2007.

\bibitem[Grisoni et~al.(2014)Grisoni, Cassotti, and Todeschini]{Grisoni2014}
F.~Grisoni, M.~Cassotti, and R.~Todeschini.
\newblock Reshaped sequential replacement for variable selection in {QSPR}:
  comparison with other reference methods.
\newblock \emph{J. Chemom.}, 28\penalty0 (4):\penalty0 249--259, 2014.
\newblock ISSN 1099-128X.

\bibitem[Hastie et~al.(2009)Hastie, Tibshirani, and Friedman]{hastie2009}
T.~Hastie, R.~Tibshirani, and J.~Friedman.
\newblock \emph{The Elements of Statistical Learning: Data Mining, Inference,
  and Prediction}.
\newblock Springer Verlag, New York, 2nd edition, 2009.

\bibitem[Leardi(2007)]{leardi2007}
R.~Leardi.
\newblock Genetic algorithms in chemistry.
\newblock \emph{J. Chromatogr. A}, 1158:\penalty0 226--233, 2007.

\bibitem[Leardi and González(1998)]{leardi1998}
R.~Leardi and A.~L. González.
\newblock Genetic algorithms applied to feature selection in {PLS} regression:
  how and when to use them.
\newblock \emph{Chemom. Intell. Lab. Syst.}, 41\penalty0 (2):\penalty0
  195--207, 1998.
\newblock ISSN 0169-7439.

\bibitem[Niazi and Leardi(2012)]{Niazi2012}
A.~Niazi and R.~Leardi.
\newblock Genetic algorithms in chemometrics.
\newblock \emph{J. Chemom.}, 26\penalty0 (6):\penalty0 345--351, 2012.
\newblock ISSN 1099-128X.

\bibitem[{R Core Team}(2013)]{rsoftware}
{R Core Team}.
\newblock \emph{R: A Language and Environment for Statistical Computing}.
\newblock R Foundation for Statistical Computing, Vienna, Austria, 2013.

\bibitem[Thomasson and Kapadia(1968)]{Thomasson1968}
R.~Thomasson and C.~Kapadia.
\newblock On estimating the parameter of a truncated geometric distribution.
\newblock \emph{Ann. Inst. Statist. Math.}, 20\penalty0 (1):\penalty0 519--523,
  1968.
\newblock ISSN 0020-3157.

\bibitem[Varmuza and Filzmoser(2009)]{Varmuza2009}
K.~Varmuza and P.~Filzmoser.
\newblock \emph{Introduction to Multivariate Statistical Analysis in
  Chemometrics}.
\newblock CRC Press, Boca Raton, FL, 2009.
\newblock ISBN 978-1-4200-5949-6.

\bibitem[Varmuza et~al.(2013)Varmuza, Filzmoser, and Dehmer]{varmuza2013}
K.~Varmuza, P.~Filzmoser, and M.~Dehmer.
\newblock Multivariate linear {QSPR/QSAR} models: Rigorous evaluation of
  variable selection for {PLS}.
\newblock \emph{Comput. Struct. Biotechnol. J.}, 5: e201302007\penalty0
  (6):\penalty0 1--10, February 2013.

\bibitem[Wehrens(2011)]{wehrens2011}
R.~Wehrens.
\newblock \emph{Chemometrics with R}.
\newblock Use R! Springer, Berlin Heidelberg, 2011.

\bibitem[Westad and Martens(2000)]{Westad2000}
F.~Westad and H.~Martens.
\newblock Variable selection in near infrared spectroscopy based on
  significance testing in partial least squares regression.
\newblock \emph{J. Near Infrared Spectrosc.}, 8\penalty0 (2):\penalty0
  117--124, 2000.

\bibitem[Wold et~al.(2001)Wold, Sj\"ostr\"om, and Eriksson]{Wold2001}
S.~Wold, M.~Sj\"ostr\"om, and L.~Eriksson.
\newblock {PLS}-regression: a basic tool of chemometrics.
\newblock \emph{Chemom. Intell. Lab. Syst.}, 58\penalty0 (2):\penalty0 109 --
  130, 2001.
\newblock ISSN 0169-7439.

\end{thebibliography}
\end{document}